\documentclass[nofootinbib,aps,amsmath,amssymb,twocolumn,pra,superscriptaddress]{revtex4-2}

\usepackage{graphicx}
\usepackage{dcolumn}
\usepackage{bm}
\usepackage{epsfig,hyperref,amsthm}
\usepackage{mathtools}
\usepackage{color}
\usepackage{colordvi}
\usepackage[dvipsnames]{xcolor}
\usepackage{relsize}
\usepackage{accents}
\usepackage{wasysym}  
\usepackage{xypic}
\usepackage{times}
\usepackage{newtxmath}

\hypersetup{
	colorlinks=true,
	linkcolor=CitingColor,
	citecolor=CitingColor,
	urlcolor=CitingColor
}

\DeclareMathAlphabet\mathbfcal{OMS}{cmsy}{b}{n}
\newcommand{\ket}[1]{\ensuremath{|#1\rangle}}
\newcommand{\bra}[1]{\ensuremath{\langle #1|}}

\newcommand{\proj}[1]{\ket{#1}\bra{#1}}
\newcommand{\be}{\begin{equation}}
\newcommand{\ee}{\end{equation}}
\newcommand{\ba}{\begin{eqnarray}}
\newcommand{\ea}{\end{eqnarray}}

\newcommand{\norm}[1]{\left\|#1\right\|}

\newcommand{\id}{\mathbb{I}}

\newtheorem{result}{Result}
\newtheorem{result-coro}[result]{\blu Corollary}

\newtheorem{question}{Question}

\definecolor{nred}{rgb}{0.9,0.1,0.1}
\definecolor{nblack}{rgb}{0,0,0}
\definecolor{nblue}{rgb}{0.2,0.2,0.8}
\definecolor{ngreen}{rgb}{0.2,0.6,0.2}
\definecolor{ublue}{rgb}{0,0,0.5}

\definecolor{pur}{rgb}{0.75,0,0.75}
\definecolor{nngrn}{rgb}{0,0.5,0.5}
\definecolor{CitingColor}{rgb}{0,0.3,1}

\newcommand{\blu}{\color{nblue}}

\newcommand{\CY}[1]{{\color{black}#1}}
\newcommand{\valerio}[1]{{\color{black}#1}}
\newcommand{\CYtwo}[1]{{\color{black}#1}}
\newcommand{\ben}[1]{{\color{black}#1}}

\newcommand{\CYnew}[1]{{\color{black}#1}}

\begin{document}
\title{
Informational non-equilibrium concentration}

\author{Chung-Yun Hsieh}
\affiliation{H. H. Wills Physics Laboratory, University of Bristol, Tyndall Avenue, Bristol, BS8 1TL, UK}

\author{Benjamin Stratton}
\affiliation{Quantum Engineering Centre for Doctoral Training, H. H. Wills Physics Laboratory and Department of Electrical \& Electronic Engineering, University of Bristol, BS8 1FD, UK}
\affiliation{H. H. Wills Physics Laboratory, University of Bristol, Tyndall Avenue, Bristol, BS8 1TL, UK}

\author{Hao-Cheng Weng}
\affiliation{Quantum Engineering Technology Laboratories and H. H. Wills Physics Laboratory, University of Bristol, BS8 1UB, UK}

\author{Valerio Scarani}
\email{physv@nus.edu.sg}
\affiliation{Centre for Quantum Technologies, National University of Singapore, 3 Science Drive 2, Singapore 117543}
\affiliation{Department of Physics, National University of Singapore, 2 Science Drive 3, Singapore 117542}

\date{\today}

\begin{abstract}
Informational contributions to thermodynamics can be studied in isolation by considering systems with fully-degenerate Hamiltonians. In this regime, being in non-equilibrium---termed {\em informational non-equilibrium}---provides thermodynamic resources, such as extractable work, {\em solely} from the information content.
The usefulness of informational non-equilibrium creates an incentive to obtain more of it, motivating the question of how to {\em concentrate it}: can we increase the local informational non-equilibrium of a product state $\rho\otimes\rho$ under a global closed system (unitary) evolution? 
We fully solve this problem analytically, showing that it is {\em impossible} for two-qubits, and it is always possible to find states achieving this in higher dimensions.
\CY{\ben{Specifically f}or two-qutrits, \ben{we find that there} is a single unitary achieving optimal concentration for {\em every} state, \ben{for which we} uncover a {\em Mepmba-like effect}. We further discuss} the notion of {\em bound resources} in this framework, initial global correlations' ability to {\em activate} concentration, \CYtwo{and applications to concentrating} purity and intrinsic randomness.  
\end{abstract}

\maketitle

\section{Introduction}
Information is central to our modern understanding of thermodynamics~\cite{Parrondo2015}.
To model a system's thermodynamic behaviours, one must consider both its energy and information contents~\cite{Sparaciari_2020}.
In fact, control over one allows influence over the other---by manipulating a system's information content, one can cool the system down via algorithmic cooling~\cite{Boykin_2002}, convert bits into work via Szilard engine~\cite{Szilard1929}, or transmit energy~\cite{Hsieh2025PRL,Hsieh2025PRA}. Alternately, by consuming energy, one can manipulate encoded information, e.g., by erasing information via Landauer's principle~\cite{Landauer1961} or performing computation~\cite{conte2019thermodynamiccomputing, aifer2024thermodynamiclinearalgebra, lipkabartosik2023thermodynamiccomputingautonomousquantum}.  

The informational contributions to thermodynamics can be isolated from the energetic ones---allowing them to be independently studied and quantified---by considering fully-degenerate Hamiltonians~\cite{Purity-review}. In the absence of energy gaps, thermodynamic transformations must arise from information processing. In this regime, thermal equilibrium is described by the maximally mixed state, and all other states are considered to be in \textit{informational non-equilibrium}. This notion coincides with purity when considering a fixed system size~\cite{Purity-review}, allowing purity also to be studied within this framework. By understanding this special case of thermodynamics, insights can be gained into the general case, where both energy and information are considered. 

Given that informational non-equilibrium (and hence purity) is a resource in thermodynamics~\cite{ChitambarRMP2019,Purity-review,Streltsov2018NJP,PhysRevA.71.062307,PhysRevA.67.062104,Stratton2024PRL}, it is natural to want to increase the amount one has. Such questions have previously been considered via \CY{\em resource distillation}~\cite{ChitambarRMP2019}, where \CY{several} copies of a less resourceful state are converted into fewer copies of a more resourceful state, with the help of an arbitrary (possibly infinite) supply of free states. 
\ben{
\CYtwo{Additionally, these protocols} may also require global operations that act simultaneously on many copies of the system and free states. Practically implementing resource distillation can, therefore, be highly expensive \CYtwo{in labs}, well beyond the ability of existing quantum technological platforms (e.g., \CYtwo{Refs.~\cite{dolde2013room, dolde2014high,bradley2019ten})}.}
\ben{As being able to concentrate multiple noisy objects into a single, more resourceful object is crucial for quantum technologies, we here focus on the {\em smallest}---and hence {\em most practically feasible}---setting for which resource enhancement can be studied. We coin the term {\em resource concentration} to describe this paradigm:}
given two copies of a state, $\rho_A \otimes \rho_B$, can we enhance the informational non-equilibrium in $A$ via a global unitary?  
\ben{The aim is thus to concentrate as much of the resource as possible locally, using only globally resource preserving operations.}
\ben{Note, we have defined the task without access to any free states: it is a closed system dynamics, which will also allow us to keep a complete accounting of the information changes. This differs from resource distillation protocols that only concern the input and output states, ignoring any ``junk'' produced in the process.}

\CY{Here,} we fully solve the resource concentration problem for informational non-equilibrium and purity and further investigate the concentration of intrinsic randomness~\cite{Meng2023}. 
\CY{Unexpectedly, our framework uncovers a phenomenon in resource concentration that is similar to Mpemba effect~\cite{Mpemba1969,Strachan2024,Rylands2024PRL,Joshi2024PRL,Aharony_Shapira2024PRL}.}

\CYnew{\section{Purity and informational non-equilibrium}}
Consider a quantum system with dimension $d<\infty$.
\CY{Qualitatively, {\em purity} of this system is a physical property about whether it is in a pure state. 
On the other hand, {\em informational non-equilibrium} of this system is another physical property about whether it is {\em not} in the maximally mixed state $\id^{(d)}/d$, where $\id^{(d)}$ is the identity operator [the superscript ``$(d)$'' denotes the dimension dependence whenever needed]. 
Hence, purity is {\em independent} of the actual physical dimension $d$, while informational non-equilibrium is, by definition, {\em dependent} on $d$. 
It is thus clear that informational non-equilibrium and purity are two {\em different} properties.}
\CY{In a more quantitative language, if the system is in a state $\rho$, then its purity captures how close $\rho$ is to some pure state, while its informational non-equilibrium quantifies how distant $\rho$ is from $\id^{(d)}/d$.}
As an example, as well known, most \CY{practically realisible} ``qubits'' are actually two levels of a multilevel system. The state $\id^{(2)}/2$ of those two levels is not a resource \CY{if} one stays in that subspace, but becomes a resource if one starts accessing other levels. Its purity is, of course, the same. Since our aim is to study how the informational non-equilibrium can be increased, we have to steer clear of the trivial way that consists in just redefining the dimension. In all that follows, the dimension is fixed, and we aim at increasing the resource by quantum operations on \CY{states.}

\subsection{Quantifying informational non-equilibrium}
Before stating our central question, we need to {\em quantify} informational non-equilibrium.
To this end, for a $d$-dimensional state $\rho$, we adopt the following figure-of-merit:
\begin{align}\label{Eq:measure}
\mathcal{P}(\rho)\coloneqq D_{\rm max}(\rho\,\|\,\id^{(d)}/d).
\end{align}
Here, 
$
D_{\rm max}(\rho\,\|\,\sigma)\coloneqq\log_2\min\{\lambda\ge0\,|\,\rho\le\lambda\sigma\}
$
is the {\em max-relative entropy}~\cite{Datta2009}, widely used for its numerical feasibility and \CYtwo{operational relevance~~\cite{Takagi2019,Hsieh2024Quantum,Hsieh2021PRXQ,Takagi2020PRL}.}
Explicitly, 
\begin{align}\label{Eq:useful}
\mathcal{P}(\rho) = \log_2d\norm{\rho}_\infty.
\end{align}
Hence, $\mathcal{P}$ quantifies informational non-equilibrium by checking $\rho$'s most ``non-maximally-mixed'' eigenvalue.
\CYtwo{In general,} $\mathcal{P}$ can also \CYtwo{act}
as a dimension-{\em dependent} measure of purity. When the system is a qubit, up to a unitary, any state reads \mbox{$\rho = \norm{\rho}_\infty\proj{0} + (1-\norm{\rho}_\infty)\proj{1}$} with $\norm{\rho}_\infty\geq 1/2$. Thus, any non-decreasing function of $\norm{\rho}_\infty$ can 
\CYtwo{quantify purity.}

\CYnew{\section{Informational Non-Equilibrium Concentration Problems}}

\subsection{Defining the concentration problems}
\begin{figure}
\includegraphics[width=3.5cm]{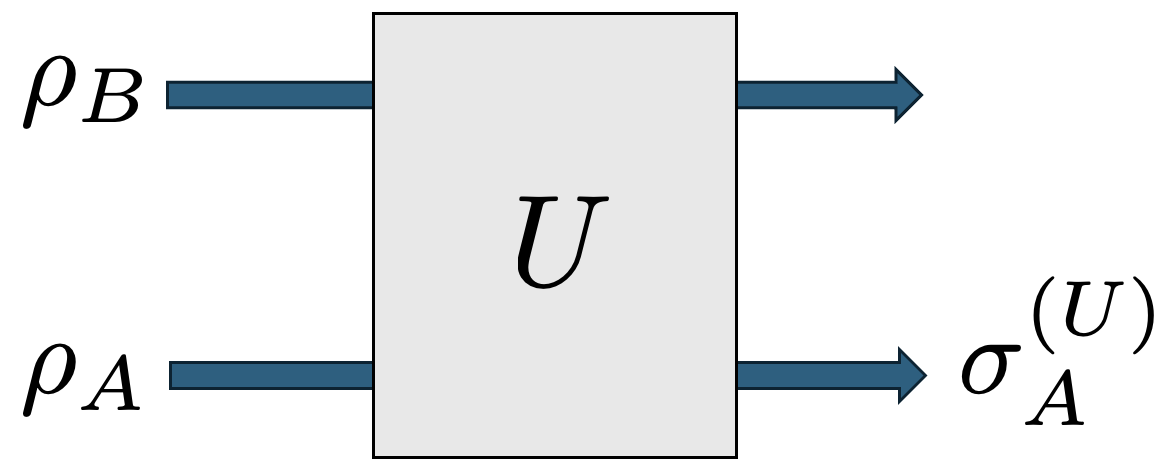}
\caption{
{\bf Informational non-equilibrium concentration problems.}
For two copies of a state $\rho$, we study whether one can use a global unitary $U_{AB}$ to enhance the informational non-equilibrium locally in $A$, in the sense that $\mathcal{P}\left(\sigma_A^{(U)}\right)>\mathcal{P}(\rho_A)$.
[$\mathcal{P}$ is defined in Eq.~\eqref{Eq:measure}].
}
\label{Fig:main problem}
\end{figure}

Now, we can define {\em informational non-equilibrium concentration problems} (INCPs).
For a $d$-dimensional state $\rho$, its INCP asks: {\em is there a two-qudit unitary $U_{AB}$ achieving}
$
\mathcal{P}\left(\sigma_A^{(U)}\right)>\mathcal{P}(\rho_A),
$
{\em where}
$
\sigma_A^{(U)}\coloneqq{\rm tr}_B[U_{AB}(\rho_A\otimes\rho_B)U_{AB}^\dagger]?
$
See also Fig.~\ref{Fig:main problem}.
Namely, can one use a closed system operation in the two-qudit system to concentrate informational non-equilibrium into the first system ($A$)?
When this is possible, we say the unitary $U$ is a solution to the state-dependent INCP of the state $\rho$.
\CYtwo{Note, in
this work,} subscripts denote the subsystems 
\CYtwo{the operators live.}

Here, we only consider closed system dynamics (i.e., unitary), rather than channels (i.e., completely-positive trace-preserving linear maps~\cite{QIC-book}). In addition to allowing for detailed accounting of information changes in the system, this prevents situations in which a channel could discard a state and replace it with a pure one (i.e., using the environment as a ``purity bank''). Moreover, we only assess the ability to concentrate informational non-equilibrium when given two copies of the {\em same} state. This is the simplest instance of an INCP, and relaxation of this restriction is left for future work.
As a remark, INCPs are related to (but different from) algorithmic cooling~\cite{Boykin_2002}.
More precisely, INCPs can be considered as a specific form of algorithmic cooling in which both the target system and machine are initially in the same state and only unitary evolution is allowed to achieve the cooling. Applying these restrictions allows for a complete analytical solution to the optimal algorithmic cooling protocol to be found when considering the figure-of-merit defined in Eq.~\eqref{Eq:measure}.

To solve INCPs, we first present a result including INCPs as special cases.
With a given bipartite system $AB$ with (not necessarily equal) local dimensions $d_A,d_B$, we define the map:
\begin{align}
\mathcal{E}_{\left(U_{AB},\eta_B\right)}\left(\rho_A\right)\coloneqq{\rm tr}_B\left[U_{AB}(\rho_A\otimes\eta_B)U_{AB}^\dagger\right],
\end{align}
where $\rho_A$ ($\eta_B$) is with dimension $d_A$ ($d_B$), and $U_{AB}$ is a unitary acting on $AB$.
Then, in \CYnew{Appendix~\ref{App:Proof of Result 1}}, we show that
\begin{result}\label{Result}
Given $d_A,d_B$, then, for every $U_{AB},\rho_A,\eta_B$, we have
\begin{align}
\max_{U_{AB}}2^{\mathcal{P}\left[\mathcal{E}_{\left(U_{AB},\eta_B\right)}\left(\rho_A\right)\right]} = \max_{\Pi^{(d_B)}_{AB}}d_A{\rm tr}\left[\Pi^{(d_B)}_{AB}\left(\rho_A\otimes\eta_B\right)\right].
\end{align}
``$\max_{\Pi^{(d_B)}_{AB}}$'' maximises over all rank-$d_B$ projector in $AB$.
\end{result}

\subsection{Solving informational non-equilibrium concentration problems}
Result~\ref{Result} fully quantifies the optimal performance of relocating informational non-equilibrium from $B$ to $A$. 
We can solve an INCP by computing the following difference:
\begin{align}
\Delta\mathcal{P}(\rho)&\coloneqq\max_{U_{AB}}\mathcal{P}\left[\mathcal{E}_{\left(U_{AB},\rho_B\right)}\left(\rho_A\right)\right] - \mathcal{P}(\rho_A)\nonumber\\
&= \max_{\Pi^{(d_B)}_{AB}}\log_2\left({\rm tr}\left[\Pi^{(d_B)}_{AB}\left(\rho_A\otimes\rho_B\right)\right]/\norm{\rho_A}_\infty\right),
\end{align}
where we have set $\eta = \rho$ in Result~\ref{Result}.
By solving the above optimisation, once the optimal value is positive, informational non-equilibrium can be concentrated in $A$ with the initial state $\rho_A\otimes\rho_B$; namely, $\rho$'s INCP has a solution. 
To further solve this, let us write $\rho = \sum_{i=0}^{d-1}a_i^\downarrow\proj{i}$, where $d=d_A=d_B$ and $a_i^\downarrow\ge a_{i+1}^\downarrow$ for every $i$.
Then we have
\begin{align}
\max_{\Pi^{(d)}_{AB}}{\rm tr}\left[\Pi^{(d)}_{AB}\left(\rho_A\otimes\rho_B\right)\right] = \max_{\Pi^{(d)}_{AB}}\sum_{ij}a_i^\downarrow a_j^\downarrow\bra{ij}\Pi^{(d)}_{AB}\ket{ij}.
\end{align}
Let us order the sequence $\{a_i^\downarrow a_j^\downarrow\}_{i,j=0}^{d-1}$ again in a non-increasing way, and let us call the re-ordered sequence $\{c_k^\downarrow(\rho)\}_{k=0}^{d^2-1}$; namely, for every $k$, we have $c_k^\downarrow(\rho) = a_i^\downarrow a_j^\downarrow$ for some $i,j$ such that each pair $(i,j)$ appears exactly once, and  $c_k^\downarrow(\rho)\ge c_{k+1}^\downarrow(\rho)$. 
Physically, $\{c_k^\downarrow(\rho)\}_{k=0}^{d^2-1}$ is the set of ordered eigenvalues of $\rho\otimes\rho$.
Finally, for a normal operator $M$, its \mbox{\em Ky Fan $K$-norm}~\cite{Fan1951}, $\norm{M}_{\text{$K$-KF}}$, is defined as the sum of its $K$ largest eigenvalues.
With this notion, we obtain
\begin{align}\label{Eq:d-KF}
\max_{\Pi^{(d)}_{AB}}{\rm tr}\left[\Pi^{(d)}_{AB}\left(\rho_A\otimes\rho_B\right)\right]=\sum_{k=0}^{d-1}c_k^\downarrow(\rho)=\norm{\rho\otimes\rho}_{\text{$d$-KF}};
\end{align}
i.e., it is the Ky Fan $d$-norm of $\rho\otimes\rho$.
Then, we arrive at the following analytical expression, serving as the complete solution to any finite-dimensional INCP:
\begin{result}\label{Result:PCP_full_solution}
For a $d$-dimensional state $\rho$, we have
\begin{align}
\Delta\mathcal{P}(\rho)= \log_2\left(\norm{\rho\otimes\rho}_{\text{$d$-{\rm KF}}}/\norm{\rho}_\infty\right).
\end{align}
\CYtwo{$\rho$'s INCP} has a solution if and only if
\mbox{$
\norm{\rho\otimes\rho}_{\text{$d$-{\rm KF}}} > \norm{\rho}_\infty.
$}
\end{result}

The Ky Fan norm has previously been used to bound the ability of thermal operations~\cite{Horodecki2013} to cool systems~\cite{Theurer_2023}. Result~\ref{Result:PCP_full_solution} now provides it with a novel operational meaning---it quantifies the optimal amount of informational non-equilibrium (and also purity) that can be concentrated given two copies of a state via unitary dynamics. Moreover, as well as providing an analytical necessary and sufficient condition for the existence of INCPs' solutions, Result~\ref{Result:PCP_full_solution} also tells us the fundamental limitation of purity concentration; i.e., $\Delta\mathcal{P}(\rho)$ is the highest \CYtwo{concentratable} amount 
with a fixed dimension.

\subsection{No two-qubit concentration of informational non-equilibrium and purity}
It is rather surprising to know that we (only) cannot concentrate informational non-equilibrium and purity in the simplest case---two-qubits.
Before stating the result, we recall that, as we have argued before, for a qubit state $\rho$, increasing purity {\em is equivalent to} enhancing $\norm{\rho}_\infty$.
Then, in \CYnew{Appendix~\ref{App:Proof of Result 3}}, we prove the following no-go result:
\begin{result}\label{Result:qubit no-go}
INCPs of qubit states have no solution.
Moreover, this conclusion is independent of the choice of purity measure.
\end{result}
Hence, for two qubits, the structure of quantum theory forbids any possible concentration of informational non-equilibrium and purity.
Moreover, this fundamental limitation is true {\em independent} of the measure that we use.

\subsection{Informational non-equilibrium concentration beyond qubits is possible}
It turns out that informational non-equilibrium concentration is a {\em generic} phenomenon existing beyond qubits.
This is because the necessary and sufficient condition for INCP's solutions to exist (Result~\ref{Result:PCP_full_solution}) can always be satisfied by some $\rho$ when the local dimension $d$ is strictly greater than $2$.
To better illustrate this, let us consider a simple example, which is an {\em effective qubit} in a qudit: \mbox{$\rho = p\proj{0} + (1-p)\proj{1}$} with \mbox{$1/2\le p\le1$} in a $d$-dimensional system with $d>2$.
As long as $p<1$, we have 
\begin{align}
\CYnew{\norm{\rho\otimes\rho}_{\text{$d$-{\rm KF}}}\ge p^2 + 2p(1-p) > \norm{\rho}_\infty,}
\end{align}
which implies concentration due to Result~\ref{Result:PCP_full_solution}.
This means that concentration of informational non-equilibrium and purity can indeed happen. 
In fact, the following state-{\em independent} unitary can do the job:
\begin{align}\label{Eq:simple unitary}
U^{\rm simple}_{AB}:\ket{10}_{AB}\leftrightarrow\ket{02}_{AB}.
\end{align}
Hence, when the local system is beyond a single qubit, INCPs, in general, can have solutions. 
Finally, we note a simple upper bound 
\begin{align}\label{Eq:Delta P < P}
\CYnew{\Delta\mathcal{P}(\rho)\le\mathcal{P}(\rho),}
\end{align}
which means that the {\em initial} informational non-equilibrium limits the optimal concentration. See \CYnew{Appendix~\ref{App:Eq:DeltaP}} for proof.

\begin{figure*}
	\includegraphics[scale=0.36]{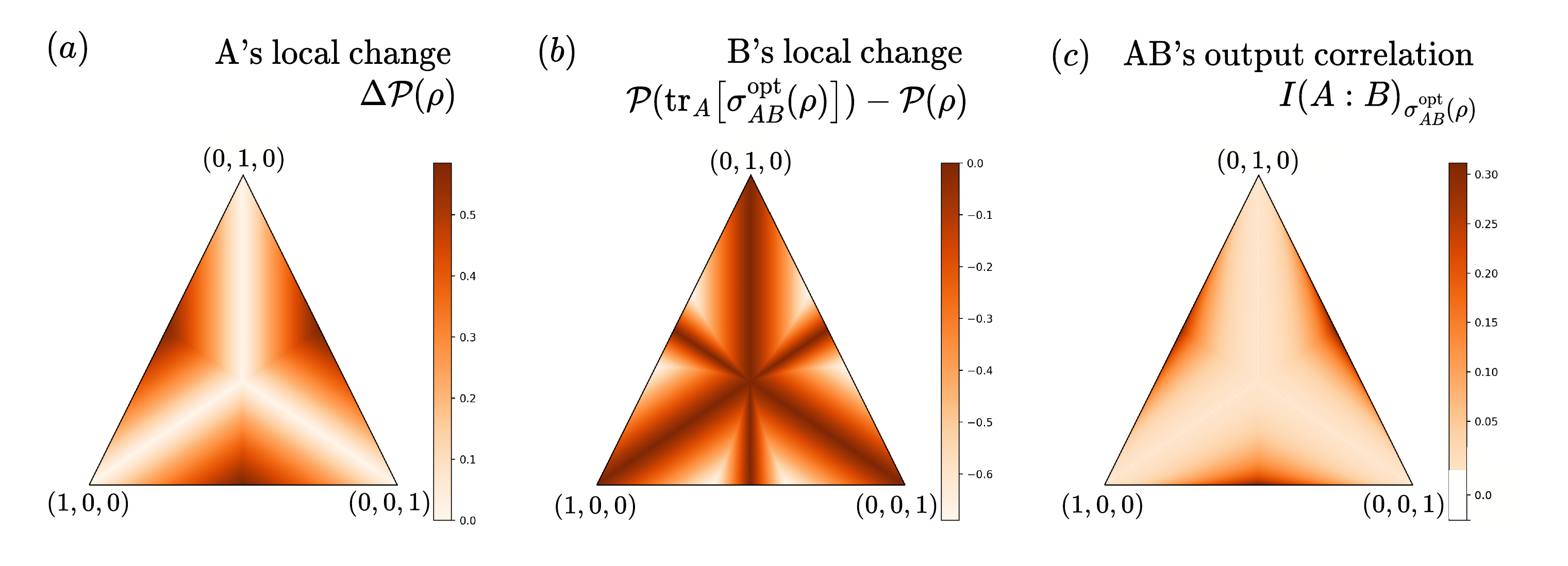}
	\caption{
 {\bf Graphical depictions of two-qutrit cases.}
 Here, we plot the analytical result Eq.~\eqref{Eq:qutrit result}.
 Each point in the triangle, $(a_0, a_1, a_2)$, represents the eigenvalues of the qutrit state, $\rho$, with the colour giving the change of informational non-equilibrium.
 (a) Optimal increment in $A$ according to Eq.~\eqref{Eq:qutrit result}. States with bound resources are those on the white lines running from the corners to the centre of the triangle. 
 (b) Change in $B$ when $A$ achieves the optimal increment $\Delta\mathcal{P}(\rho)$.
(c) The mutual information between $A$ and $B$ after the optimal concentration. One can then see that $\Delta\mathcal{P}>0$ is accompanied with non-vanishing correlation, as claimed in Result~\ref{result:corr}. States for which no correlations are created have been explicitly highlighted in white, and can be seen to coincide with the states possessing bound purity.
	\label{Fig:example}}
	\end{figure*}

\subsection{Optimal two-qutrit purity concentration must generate global correlations}
Now, we know concentrating purity is possible via INCPs.
A natural question is: {\em can we concentrate local purity without generating global correlation?}
In the two-qutrit case, we show that, surprisingly, it is {\em impossible} due to the special structure of qutrits. 
More precisely, the three largest eigenvalues of a two-qutrit state $\rho_A\otimes\rho_B$ are $a_0^\downarrow a_0^\downarrow, a_0^\downarrow a_1^\downarrow,a_1^\downarrow a_0^\downarrow$.
Then, Result~\ref{Result:PCP_full_solution} implies that
\begin{align}\label{Eq:qutrit result}
\Delta\mathcal{P}(\rho) = \log_2\left(a_0^\downarrow+2a_1^\downarrow\right).
\end{align}
For a better understanding, we plot the analytical result Eq.~\eqref{Eq:qutrit result} in Fig.~\ref{Fig:example}.
Also, one can check that the unitary $U^{\rm simple}_{AB}$ defined in Eq.~\eqref{Eq:simple unitary} achieves Eq.~\eqref{Eq:qutrit result}; i.e., $U^{\rm simple}_{AB}$ is {\em optimal}.
Let \mbox{$\sigma^{\rm opt}_{AB}(\rho) \coloneqq U^{\rm simple}_{AB}(\rho_A\otimes\rho_B)U^{\rm simple,\dagger}_{AB}$} be $U^{\rm simple}_{AB}$'s global output.
To quantify global output's correlation, we use the quantum mutual information, a widely-used correlation measure.
Formally, for a bipartite state $\eta_{AB}$, its {\em quantum mutual information}~\cite{QIC-book} is \mbox{$I(A:B)_{\eta_{AB}}\coloneqq S(\eta_A)+S(\eta_B)-S(\eta_{AB})$}~\footnote{Here, $\eta_A\coloneqq{\rm tr}_B(\eta_{AB})$ and $\eta_B\coloneqq{\rm tr}_A(\eta_{AB})$.}, where \mbox{$S(\eta)\coloneqq-{\rm tr}(\eta\log_2\eta)$} is the {\em von-Neumann entropy}~\cite{QIC-book}.
Then, \CYnew{in Appendix~\ref{App:Proof of Result 4}}, we show that
\begin{result}\label{result:corr}
Two-qutrit optimal purity concentration must generate global correlation: $I(A:B)_{\sigma^{\rm opt}_{AB}(\rho)}>0$ if $\Delta\mathcal{P}(\rho)>0$.
\end{result}
Hence, counter-intuitively, one {\em must} increase global correlation and local purity simultaneously.
Since a pure state cannot be correlated with any other system, this result means that {\em it is impossible to map a non-pure qutrit state to a perfect pure state} in the current setting. 
This finding further uncovers a trade-off relation between making local states purer and generating global correlation (and makes local states less pure).

\CYnew{\section{Physical Implications}}

\subsection{A ``Mpemba-like'' effect for purity concentration}
Equation~\eqref{Eq:qutrit result} implies that, when we optimally concentrate purity \CYtwo{in $A$,}
\CYtwo{$B$'s local purity,}
as measured by $\mathcal{P}$, can be {\em invariant}. To see this, consider the one-parameter family of qutrit states
\begin{align}
\CYnew{\rho^{(p)} = p\proj{0}+(1-p)\proj{1}}
\end{align}
with $1/2\le p\le1$.
Letting \mbox{$\rho^{(p)}\otimes\rho^{(p)}$} evolve under the optimal unitary $U_{AB}^{\rm simple}$ [Eq.~\eqref{Eq:simple unitary}], \CYtwo{$B$'s local output} reads
\begin{align}
\sigma_B^{\rm opt}\left(\rho^{(p)}\right) = p^2\proj{0} + (1-p)\proj{1}+p(1-p)\proj{2},
\end{align}
meaning that $B$ has \CY{output purity} 
\begin{align}
\mathcal{P}\left[\sigma_B^{\rm opt}\left(\rho^{(p)}\right)\right] = \log_2\left(3{\rm max}\{p^2;1-p\}\right).
\end{align}
\CY{Now, by setting $p=1/2$, Eq.~\eqref{Eq:qutrit result} says that the optimal purity increment in $A$ is $\Delta\mathcal{P}\left(\rho^{(1/2)}\right) = \log_2(3/2)>0$.
Meanwhile, locally in $B$, we have 
\begin{align}
\mathcal{P}\left[\sigma_B^{\rm opt}\left(\rho^{(1/2)}\right)\right] = \log_2(3/2) = \mathcal{P}\left(\rho^{(1/2)}\right)
\end{align}
[see also Fig.~\ref{Fig:example} and Eq.~\eqref{Eq: sigma opt B} in \CYnew{Appendix~\ref{App:Proof of Result 4}}]. Hence, purity, as measured by $\mathcal{P}$, does not change in $B$ when we optimally increase it in $A$~\footnote{\CY{Physically, this is because $\mathcal{P}$ only focuses on the ``purest'' occupation (i.e., the maximal eigenvalue). Manipulating less pure occupations cannot change $\mathcal{P}$'s value.}}---the sum of local resources is not conserved.}

\valerio{
More intriguingly, our calculation uncovers a phenomenon similar to the {\em Mpemba effect}~\cite{Mpemba1969,Strachan2024,Rylands2024PRL,Joshi2024PRL,Aharony_Shapira2024PRL}. Loosely speaking, the Mpemba effect describes the phenomenon that, evolving under a given dynamics for a fixed amount of time, an initially colder system may reach a higher final temperature than an initially hotter system (and vice versa). Here, under the dynamics $U_{AB}^{\rm simple}$, we observe that in system $B$ a {\em purer} initial state can be mapped to a {\em more mixed} final state. As an example of this, consider $p = p_+\coloneqq(\sqrt{5}-1)/2$. Clearly $\rho^{(p_+)}$ is {\em purer} than $\rho^{(1/2)}$. 
Now, when $\rho^{(p_+)}\otimes\rho^{(p_+)}$ undergoes $U^{\rm simple}_{AB}$, $B$'s local output $\sigma_B^{\rm opt}\left(\rho^{(p_+)}\right)$ satisfies
\begin{align}
\mathcal{P}\left[\sigma_B^{\rm opt}\left(\rho^{(p_+)}\right)\right]
<
\mathcal{P}\left[\sigma_B^{\rm opt}\left(\rho^{(1/2)}\right)\right].
\end{align}
So a purer environment interacting with a purer state is left in a more mixed state than an initially more mixed environment interacting with a more mixed state.} 
\CY{Intuitively, 
one might expect that if the environment ($B$) is initially purer, then optimal concentration in $A$ would leave the environment less mixed. However, here we see the {\em opposite}, with purer initial states leading to more mixed environments---a phenomenon captures a similar flavor to the Mpemba effect. We leave further explorations for future projects.}

\subsection{A notion of ``bound'' informational non-equilibrium}
From Result~\ref{Result:qubit no-go}, if a non-pure qubit state is not maximally mixed, it carries non-vanishing resources that are {\em not yet} the highest but {\em cannot} be concentrated further.
We coin the term {\em bound informational non-equilibrium} for such states, and we briefly discuss their properties here beyond qubit.  
First, in qutrits, Eq.~\eqref{Eq:qutrit result} implies that $\Delta\mathcal{P}(\rho)= 0$ {\em if and only if} \mbox{$a_1^\downarrow = a_2^\downarrow$}.
Namely, {\em all non-pure qutrit states with exactly two-fold degeneracy in their smaller eigenvalue have bound informational non-equilibrium.}
Notably, in qutrits, small perturbations are enough to remove bound informational non-equilibrium by breaking the equality $a_1^\downarrow = a_2^\downarrow$.
Meanwhile, in qubits, no perturbation can do so.
Hence, interestingly, depending on the physical system's dimension, bound informational non-equilibrium could be either very robust (when $d=2$) or very fragile (when $d=3$) against noises.
Now, generally, for a $d$-dimensional state $\rho$, Result~\ref{Result:PCP_full_solution} implies that $\Delta\mathcal{P}(\rho)=0$ {\em if and only if} \mbox{$\norm{\rho\otimes\rho}_{\text{$d$-{\rm KF}}}=\norm{\rho}_\infty$}.
This thus implies {\em all non-pure qudit states with exactly $(d-1)$-fold degeneracy in their smaller eigenvalue have bound informational non-equilibrium}---this is because all such states are of the form \mbox{$\rho{(p,\ket{\psi})}\coloneqq p\proj{\psi} + (1-p)\id/d$} for some pure state $\ket{\psi}$ and $0<p<1$, and one can check that $\norm{\rho{(p,\ket{\psi})}\otimes\rho{(p,\ket{\psi})}}_{\text{$d$-{\rm KF}}}=\norm{\rho{(p,\ket{\psi})}}_\infty$.
Physically, this means that dephasing process \hbox{$(\cdot) \mapsto p\mathcal{I}(\cdot) + (1-p)\textrm{tr}(\cdot)\id/d$} on pure states produces bound informational non-equilibrium as long as $0<p<1$. 
Namely, dephasing processes are strong enough to negate the possibility of concentration.

\subsection{Initial correlations can activate informational non-equilibrium concentration}
Importantly, by allowing initial correlation, even an almost-vanishing amount, can make informational non-equilibrium concentration possible. To see this, suppose one has the two-qudit {\em isotropic state}~\cite{Horodecki1999}
$
p\proj{\Phi^+}_{AB} + (1-p)\id_{AB}/d_{AB},
$
where \mbox{$\ket{\Phi^+}_{AB}\coloneqq\sum_{i=0}^{d-1}\ket{ii}_{AB}$} is maximally entangled and \mbox{$0\le p\le1$}. Locally, both systems are maximally mixed, a state for which no informational non-equilibrium can be concentrated.
However, by considering the two-qudit unitary that maps $\ket{\Phi^+}\leftrightarrow\ket{00}$, one can obtain non-maximally-mixed marginal, resulting in informational non-equilibrium concentration.
The physics is that one can consume the global correlation (even a classical, non-entangled one) to generate local purity.
Namely, we can relocate the genuinely global purity into local systems.
This also shows that the two-qubit no-go result (Result~\ref{Result:qubit no-go}) is not robust to practical noise and experimental error bars---one can consume global correlation to break it.
Notably, the same argument works for {\em arbitrary} \mbox{$\rho = \sum_ia_i\proj{i}$} by considering
$
p\proj{\rho}_{AB}+(1-p)\rho_A\otimes\rho_B,
$
where
$
\ket{\rho}_{AB}\coloneqq\sum_i\sqrt{a_i}\ket{ii}_{AB}
$
is $\rho$'s purification.
Hence, global correlations are useful resources for activating local concentrations of informational non-equilibrium and purity.

At this point, one may wonder: {\em to what extent can global entanglement enhance the concentration?}
This is, again, captured by the Ky Fan norm.
To see this, if two copies of $\rho$ are entangled via $\ket{\rho}_{AB}$, a global unitary mapping as \mbox{$\ket{\rho}_{AB}\leftrightarrow\ket{00}_{AB}$} can achieve concentration in $A$ with the increment $\Delta\mathcal{P}_{\rm corr}(\rho)\coloneqq\log_2d - \log_2d\norm{\rho}_\infty =-\log_2\norm{\rho}_\infty$.
Using Result~\ref{Result:PCP_full_solution}, the optimal concentration without any global correlation is $\Delta\mathcal{P}(\rho) = \log_2\left(\norm{\rho\otimes\rho}_{\text{$d$-{\rm KF}}}/\norm{\rho}_\infty\right)$.
Consuming $\ket{\rho}_{AB}$'s entanglement leads to the additional concentration 
\begin{align}
\Delta\mathcal{P}_{\rm corr}(\rho)-\Delta\mathcal{P}(\rho) = -\log\norm{\rho\otimes\rho}_{\text{$d$-{\rm KF}}}.
\end{align}
Thus, the Ky Fan norm not only characterises INCPs' solutions---it is also the {\em entanglement advantage} in INCPs.

\subsection{Application to concentrating intrinsic randomness}
Finally, as 
\CYtwo{Result~\ref{Result:PCP_full_solution}'s application,} we show that informational non-equilibrium concentration implies the ability to concentrate {\em intrinsic randomness}. The intrinsic randomness of a state $\rho$ is loosely speaking defined by choosing the measurement, such that even a powerful adversary has difficulty in guessing its outcomes. We refer to Ref.~\cite{Meng2023} for all the exact definitions, and just use the result of their optimisation: the intrinsic randomness of $\rho$ is given by $-\log P_{\rm guess}^*(\rho)$, with the guessing probability
\begin{align}
\CYnew{P_{\rm guess}^*(\rho) = \left({\rm tr}\sqrt{\rho}\right)^2/d.}
\end{align}
One can see that a smaller $P_{\rm guess}^*$ means a higher purity. In particular, given a pure state, there exist measurements whose outcomes can be maximally unpredictable [$P_{\rm guess}^*(\rho)=1/d$]; while the maximally mixed state has no intrinsic randomness since $P_{\rm guess}^*(\rho)=1$.

Despite being an alternative way to measure purity, we note that {\em $\mathcal{P}$ and $P_{\rm guess}^*$ do not define the same order on states.}
That is, \mbox{$\mathcal{P}(\sigma)>\mathcal{P}(\rho)$} does not necessarily imply \mbox{$P_{\rm guess}^*(\sigma)<P_{\rm guess}^*(\rho)$} (see \CYnew{Appendix~\ref{App:not the same order}} for the explicit example).
Hence, an increase in $\mathcal{P}$ does not automatically guarantee an increase in intrinsic randomness. Nonetheless, we show that whenever informational non-equilibrium can be concentrated (i.e., $\Delta\mathcal{P}>0$), it is always possible to increase intrinsic randomness as well (i.e., decreasing $P_{\rm guess}^*$):
\begin{result}\label{Result:randomness}
When $\Delta\mathcal{P}(\rho)>0$, there exists a pairwise permutation unitary $V:\ket{i,j}\leftrightarrow\ket{0,k}$ for some $i,j,k$ achieving
\begin{align}
\mathcal{P}\left(\sigma_A^{(V)}\right)>\mathcal{P}(\rho_A)
\quad\&\quad
P_{\rm guess}^*\left(\sigma_A^{(V)}\right)<P_{\rm guess}^*(\rho_A),
\end{align}
where $\sigma_A^{(V)}\coloneqq{\rm tr}_B[V_{AB}(\rho_A\otimes\rho_B)V_{AB}^\dagger]$.
\end{result}
The proof is given in \CYnew{Appendix~\ref{App:Proof of Result 5}}, leading to an explicit formula [Eq.~\eqref{Eq:enhanced randomness}] for the possible enhancement of $P_{\rm guess}^*$.

\subsection{Experimental Practicality}
Finally, we comment on INCPs' practical feasibility. INCPs' formulation allows them to be studied in {\em Nitrogen Vacancy} (NV) centre spin systems, considering the effect of partially non-degenerate qubit/qudit energy levels and finite difference between NV centre spin systems. In Fig.~\ref{Fig:main problem}, $\rho_A$ and $\rho_B$ can be two closely populated NV centres where $U_{AB}$ can be realized by the dipole–dipole interaction between NVs~\cite{dolde2013room, dolde2014high}. The system (and thus the dimension of qudit) can be selected from the electron spins, $\prescript{14}{}{\rm N}$ ($\prescript{15}{}{\rm N}$) nuclear spins, and $\prescript{13}{}{\rm C}$ nuclear spins sub-systems~\cite{dolde2014high, bradley2019ten}.
Further experimental explorations are beyond the scope of this work and are left for future research.

\section{Discussions}
\ben{Thermodynamically, \CYtwo{solving INCPs is} calculating the \CY{highest locally extractable work} 
given access to $\rho\otimes\rho$ and joint unitary operations. By performing \CY{optimal unitaries,} a local system is maximally driven from \CY{equilibrium,} hence becoming maximally thermodynamically resourceful, \CY{and one can} extract work from \CY{it via, e.g.,} Szilard engine \cite{Szilard1929}.

\CY{As important follow-ups, our resource concentration framework applies} to the more general thermodynamic setting with non-degenerate Hamiltonians via \CY{concentration of athermality}~\cite{Lostaglio2019} \CY{as well as other resources, such as unspeakable coherence~\cite{PhysRevA.94.052324} or entanglement.}
\CY{Finally, the Mpemba-like effect and the notion of bound purity are both worth exploring.}}

\section*{Acknowledgements}
We thank \CY{Antonio Ac\'in, Xueyuan Hu, Shuyang Meng, Peter Sidajaya, Paul Skrzypczyk, David J.~Strachan, and Lin Htoo Zaw} for fruitful discussions and comments.
We especially thank Paul Skrzypczyk for pointing out the connection between our findings and Ky Fan norms \CY{and both Paul Skrzypczyk as well as David J.~Strachan for fruitful discussions about Mpemba effect.}
C.-Y.~H. acknowledges support from the Royal Society through Enhanced Research Expenses (on grant NFQI), the ERC Advanced Grant (on grant FLQuant), and the Leverhulme Trust Early Career Fellowship (on grant ``Quantum complementarity: a novel resource for quantum science and technologies'' with number ECF-2024-310).
B.~S.~acknowledges support from UK EPSRC (EP/SO23607/1). 
V.~S.~is supported by the National Research Foundation, Singapore and A*STAR under its CQT Bridging Grant; 
by the National Research Foundation, Singapore, through the National Quantum Office, hosted in A*STAR, under its Centre for Quantum Technologies Funding Initiative (S24Q2d0009); 
and by the Ministry of Education, Singapore, under the Tier 2 grant ``Bayesian approach to irreversibility'' (Grant No.~MOE-T2EP50123-0002).

\appendix
\CYnew{\section{Proof of Result~\ref{Result}}\label{App:Proof of Result 1}}
\noindent\CYnew{\em Proof.}
Using Eq.~\eqref{Eq:useful}, we analyse
\begin{align}\label{Eq:Result1_comp001}
&\max_{U_{AB}}\norm{\mathcal{E}_{\left(U_{AB},\eta_B\right)}\left(\rho_A\right)}_\infty\nonumber\\
&= \max_{U_{AB},\ket{\phi}_A}{\rm tr}\left[U_{AB}^\dagger(\proj{\phi}_A\otimes\id_B)U_{AB}(\rho_A\otimes\eta_B)\right]\nonumber\\
&\le\max_{\Pi^{(d_B)}_{AB}}{\rm tr}\left[\Pi^{(d_B)}_{AB}(\rho_A\otimes\eta_B)\right],
\end{align}
where \mbox{$U_{AB}^\dagger(\proj{\phi}_A\otimes\id_B)U_{AB}$} is a rank-$d_B$ projector in $AB$ and results in the last inequality.
Now, we note that, for an arbitrarily given rank-$d_B$ projector $\Pi^{(d_B)}_{AB}$, we can write
\begin{align}
\Pi^{(d_B)}_{AB} = \sum_{n=1}^{d_B}\proj{\kappa_n}_{AB},
\end{align}
where $\left\{\ket{\kappa_n}_{AB}\right\}_{n=1}^{d_B}$ is an orthonormal set with $d_B$ many pure states.
By considering the unitary $\widetilde{U}_{AB}^\dagger$ mapping as
\begin{align}
\ket{0}_A\otimes\ket{n}_B\leftrightarrow\ket{\kappa_n}_{AB}\;\forall\,n,
\end{align}
and keeping all other basis states untouched, we obtain
\begin{align}
\widetilde{U}_{AB}^\dagger(\proj{0}_A\otimes\id_B)\widetilde{U}_{AB} = \Pi^{(d_B)}_{AB}.
\end{align}
Hence, the inequality in Eq.~\eqref{Eq:Result1_comp001} is achieved, and the desired result follows.
\hfill$\square$

\CYnew{\section{Proof of Result~\ref{Result:qubit no-go}}\label{App:Proof of Result 3}}
\noindent\CYnew{\em Proof.}
Write $\rho = p\proj{0} + (1-p)\proj{1}$ with \mbox{$1/2\le p\le1$.}
Using Result~\ref{Result:PCP_full_solution}, it suffices to check
\begin{align}
c_0^\downarrow(\rho) + c_1^\downarrow(\rho) = p^2 + p(1-p) = p = \norm{\rho}_\infty.
\end{align}
Hence, we can never have the strict inequality ``$>$.''
Result~\ref{Result:PCP_full_solution} implies that it is impossible to increase $\norm{\rho}_\infty$.
Importantly, in a qubit, this further means that increasing the difference between two eigenvalues is impossible.
Hence, two-qubit purity cannot be concentrated, {\em independent} of the choice of measures.
\hfill$\square$

\CYnew{\section{Proof of Eq.~\eqref{Eq:Delta P < P}}\label{App:Eq:DeltaP}}
\noindent\CYnew{\em Proof.}
A direct computation shows that
\begin{align}
&\Delta\mathcal{P}(\rho) = \mathcal{P}\left(\sigma_A^{(U)}\right) - \mathcal{P}(\rho_A)\nonumber\\
&=D_{\rm max}\left[{\rm tr}_B\left(U_{AB}(\rho_A\otimes\rho_B)U_{AB}^\dagger\right)\,\middle\|\,\id_A/d\right] - D_{\rm max}(\rho\,\|\,\id/d)\nonumber\\
&\le D_{\rm max}\left[\rho_A\otimes\rho_B\,\middle\|\,(\id_A\otimes\id_B)/d^2\right] - D_{\rm max}(\rho\,\|\,\id/d)\nonumber\\
&=D_{\rm max}(\rho\,\|\,\id/d) = \mathcal{P}(\rho),
\end{align}
where we have used the data-processing inequality under the channel ${\rm tr}_B\left(U_{AB}(\cdot)U_{AB}^\dagger\right)$, and the fact that $D_{\rm max}\left[\rho_A\otimes\rho_B\,\middle\|\,(\id_A\otimes\id_B)/d^2\right] = 2D_{\rm max}(\rho\,\|\,\id/d)$.
\hfill$\square$

Interestingly, by applying this bound to both $A$ and $B$, we conclude that the sum of local changes of informational non-equilibrium in $A$ and $B$ is upper bounded by $2\mathcal{P}(\rho)$.

\CYnew{\section{Proof of Result~\ref{result:corr}}\label{App:Proof of Result 4}}
\noindent\CYnew{\em Proof.}
First, we have
\begin{align}\label{Eq:AppendixIV001}
&\sigma^{\rm opt}_{A}(\rho)\coloneqq{\rm tr}_B\left[\sigma^{\rm opt}_{AB}(\rho)\right]\nonumber\\
&=a_0^\downarrow\left(a_0^\downarrow+2a_1^\downarrow\right)\proj{0} + \left[a_1^\downarrow a_1^\downarrow+\left(1-a_2^\downarrow\right)a_2^\downarrow\right]\proj{1}\nonumber\\
&\quad+ a_2^\downarrow\proj{2};\\
&\sigma^{\rm opt}_{B}(\rho)\coloneqq{\rm tr}_A\left[\sigma^{\rm opt}_{AB}(\rho)\right]\nonumber\\
&= a_0^\downarrow\left(a_0^\downarrow+2a_2^\downarrow\right)\proj{0} + a_1^\downarrow\proj{1}\nonumber\\
&\quad+ \left[a_2^\downarrow a_2^\downarrow+\left(1-a_1^\downarrow\right)a_1^\downarrow\right]\proj{2}.\label{Eq: sigma opt B}
\end{align}
Also, since $S\left[\sigma^{\rm opt}_{AB}(\rho)\right] = S(\rho\otimes\rho) = 2S(\rho)$, we have 
\begin{align}
I(A:B)_{\sigma^{\rm opt}_{AB}(\rho)}=S\left[\sigma^{\rm opt}_{A}(\rho)\right] + S\left[\sigma^{\rm opt}_{B}(\rho)\right]-2S(\rho),
\end{align}
which is strictly positive if $\Delta\mathcal{P}(\rho) = \log_2\left(a_0^\downarrow+2a_1^\downarrow\right)>0$, as shown in Fig.~\ref{Fig:example} (which provides further illustrations).
\hfill$\square$\\

\CYnew{\section{$\mathcal{P}$ and $P_{\rm guess}^*$ do not define the same order on states}\label{App:not the same order}}
To see a counterexample, in a five-level system, consider states 
\begin{align}
\CYnew{\sigma = \proj{0}/2 + (\proj{1}+\proj{2}+\proj{3}+\proj{4})/8}
\end{align}
and 
\begin{align}
\CYnew{\rho = (\proj{0}+\proj{1}+\proj{2})/3.}
\end{align}
Then we have 
\begin{align}
\mathcal{P}(\sigma)=\log_2(5/2)>\log_2(5/3)=\mathcal{P}(\rho).
\end{align}
At the same time, we also have 
\begin{align}
P_{\rm guess}^*(\sigma) = (1/\sqrt{2}+\sqrt{2})^2/5>3/5=P_{\rm guess}^*(\rho).
\end{align}
Hence, \mbox{$\mathcal{P}(\sigma)>\mathcal{P}(\rho)$} does not necessarily imply \mbox{$P_{\rm guess}^*(\sigma)<P_{\rm guess}^*(\rho)$}.\\

\CYnew{\section{Proof of Result~\ref{Result:randomness}}\label{App:Proof of Result 5}}
\noindent\CYnew{\em Proof.}
Using Result~\ref{Result:PCP_full_solution}, $\Delta\mathcal{P}(\rho)>0$ implies
\begin{align}
\sum_{k=0}^{d-1}c_k^\downarrow(\rho) > \norm{\rho}_\infty = \sum_{i=0}^{d-1}\norm{\rho}_\infty a_i^\downarrow,
\end{align}
where we recall that $\rho = \sum_{i=0}^{d-1}a_i^\downarrow\proj{i}$ and $a_i^\downarrow\ge a_{i+1}^\downarrow$ $\forall\,i$.
By construction, we must have $c_0^\downarrow(\rho) = \norm{\rho}_\infty^2$ and $a_{0}^\downarrow = \norm{\rho}_\infty$.
Consequently, we have
\begin{align}
\sum_{k=1}^{d-1}\left(c_k^\downarrow(\rho) - \norm{\rho}_\infty a_k^\downarrow\right) > 0.
\end{align} 
This means there exists at least one $k$ value, say $k_*$, achieving
\begin{align}
c_{k_*}^\downarrow(\rho) > \norm{\rho}_\infty a_{k_*}^\downarrow.
\end{align}
Let us write $c_{k_*}^\downarrow(\rho) = a_{i_*}^\downarrow a_{j_*}^\downarrow$ for some indices $i_*,j_*$.
Then the inequality $c_{k_*}^\downarrow(\rho) > \norm{\rho}_\infty a_{k_*}^\downarrow$ can be translated into
\begin{align}
a_{i_*}^\downarrow a_{j_*}^\downarrow > a_{0}^\downarrow a_{k_*}^\downarrow.
\end{align}
Now consider the pairwise permutation unitary
\begin{align}
V_{AB}: \ket{i_*,j_*}\leftrightarrow\ket{0,k_*}.
\end{align}
Define $\delta_*\coloneqq a_{i_*}^\downarrow a_{j_*}^\downarrow - a_{0}^\downarrow a_{k_*}^\downarrow > 0.$
Then, one can check that
\begin{align}\label{Eq:sigma_V_formula}
\sigma_A^{(V)}&\coloneqq{\rm tr}_B[V_{AB}(\rho_A\otimes\rho_B)V_{AB}^\dagger]\nonumber\\
&= \rho_A + \delta_*(\proj{0}_A-\proj{i_*}_A).
\end{align}
This means that $\mathcal{P}\left(\sigma_A^{(V)}\right)>\mathcal{P}(\rho_A)$, since the occupation of $\ket{0}$ is increased by $\delta_*$.
The final step is to argue that this unitary is able to decrease the guessing probability.
Since $P_{\rm guess}^*(\rho_A) = \left({\rm tr}\sqrt{\rho_A}\right)^2/d$~\cite{Meng2023}, decreasing $P_{\rm guess}^*$ is equivalent to decreasing ${\rm tr}\sqrt{\rho_A}$; namely, it suffices to show that \mbox{${\rm tr}\sqrt{\rho_A} > {\rm tr}\sqrt{\sigma^{(V)}_A}$.}
Then a direct computation shows that (remember that $a_{0}^\downarrow$ is the largest one among all $a_{i}^\downarrow$'s)
\begin{align}
&\left(\sqrt{a_{i_*}^\downarrow-\delta_*} + \sqrt{a_{i_*}^\downarrow}\right)\left(\sqrt{a_{0}^\downarrow+\delta_*} - \sqrt{a_{0}^\downarrow}\right)\nonumber\\
&<\left(\sqrt{a_{0}^\downarrow+\delta_*} + \sqrt{a_{0}^\downarrow}\right)\left(\sqrt{a_{0}^\downarrow+\delta_*} - \sqrt{a_{0}^\downarrow}\right) = \delta_*\nonumber\\
&=\left(\sqrt{a_{i_*}^\downarrow-\delta_*} + \sqrt{a_{i_*}^\downarrow}\right)\left(\sqrt{a_{i_*}^\downarrow} - \sqrt{a_{i_*}^\downarrow-\delta_*}\right).
\end{align}
Note that we have the above strict inequality because \mbox{$\delta_*>0$} and $\sqrt{a_{i_*}^\downarrow-\delta_*} + \sqrt{a_{i_*}^\downarrow}>0$ (it cannot be zero, otherwise we cannot have $\delta_*>0$). 
Hence, we conclude that
\begin{align}
\sqrt{a_{0}^\downarrow+\delta_*} - \sqrt{a_{0}^\downarrow} < \sqrt{a_{i_*}^\downarrow} - \sqrt{a_{i_*}^\downarrow-\delta_*}.
\end{align}
Finally, we note that $\sqrt{\rho_A}$ and $\sqrt{\sigma^{(V)}_A}$ are different only in the subspace spanned by $\ket{i_*}$ and $\ket{0}$.
This can be explicitly seen by Eq.~\eqref{Eq:sigma_V_formula}.
Consequently, one can check that
\begin{align}\label{Eq:enhanced randomness}
{\rm tr}\sqrt{\rho_A} - {\rm tr}\sqrt{\sigma^{(V)}_A} = \sqrt{a_{0}^\downarrow} + \sqrt{a_{i_*}^\downarrow} - \sqrt{a_{0}^\downarrow+\delta_*} - \sqrt{a_{i_*}^\downarrow-\delta_*} >0,
\end{align}
which concludes the proof.
\hfill$\square$

\bibliography{Ref.bib}

\end{document}